\documentclass[journal]{IEEEtran}
\ifCLASSOPTIONcompsoc
  \usepackage[nocompress]{cite}
\else
  \usepackage{balance}
    \usepackage{cite}
\fi
\usepackage{graphicx}
\usepackage{amsthm}
\usepackage{amsmath}
\usepackage{amssymb}
\usepackage[ruled,vlined,linesnumbered]{algorithm2e}
\usepackage{smartdiagram}
\usepackage{verbatim} 
\usepackage{lettrine}

\begin{document}
\title{Authentication and Handover Challenges and Methods for Drone Swarms}

\author {Yucel~Aydin,
        Gunes~Karabulut~Kurt,~\IEEEmembership{Senior Member, IEEE},\\
	~{Enver~Ozdemir,~\IEEEmembership{Senior Member, IEEE},} and {Halim~Yanikomeroglu,~\IEEEmembership{Fellow, IEEE}  }  
\thanks{Manuscript received January 14, 2022; accepted March 2, 2022. 

\textit{(Corresponding author: Yucel Aydin.)}}

\thanks{Yucel Aydin and Enver Ozdemir are with the Informatics Institute, Istanbul Technical University, 34485 Istanbul, Turkey (email: aydinyuc@itu.edu.tr; ozdemiren@itu.edu.tr).}

\thanks{Gunes Karabulut Kurt is with the Poly-Grames Research Center, Department of Electrical Engineering,  Polytechnique Montr\'eal, Montr\'eal, QC H3C3A7, Canada, (e-mail: gunes.kurt@polymtl.ca). }

\thanks{Halim Yanikomeroglu is with the Department of Systems and Computer Engineering, Carleton University, Ottawa, ON K1S 5B6 Canada (e-mail: halim@sce.carleton.ca).}

\thanks{Digital Object Identifier 10.1109/JRFID.2022.3158392}

%\thanks{Manuscript received ..., 2019.}
} 

\markboth{IEEE JOURNAL OF RADIO FREQUENCY IDENTIFICATION}{Authentication and Handover Challenges and Methods for Drone Swarms}
\maketitle
\begin{abstract}
Drones are begin used for various purposes such as border security, surveillance, cargo delivery, visual shows and it is not possible to overcome such intensive tasks with a single drone. In order to expedite performing such tasks, drone swarms are employed. The number of drones in a swarm can be high depending on the assigned duty. The current solution to authenticate a single drone using a 5G new radio (NR) network requires the execution of two steps. The first step covers the authentication between a drone and the 5G core network, and the second step is the authentication between the drone and the drone control station. It is not feasible to authenticate each drone in a swarm with the current solution without causing a significant latency. Authentication keys between a base station (BS) and a user equipment (UE) must be shared with the new BS while performing handover. The drone swarms are heavily mobile and require several handovers from BS to a new BS. Sharing authentication keys for each drone as explained in 5G NR is not scalable for the drone swarms. Also, the drones can be used as a UE or a radio access node on board unmanned aerial vehicle (UxNB). A UxNB may provide service to a drone swarm in a rural area or emergency. The number of handovers may increase and the process of sharing authentication keys between UxNB to new UxNB may be vulnerable to eavesdropping due to the wireless connectivity. In this work, we present a method where the time and the number of the communication for the authentication of a new drone joining the swarm are less than 5G NR. In addition, group-based handover solutions for the scenarios in which the base stations are terrestrial or mobile are proposed to overcome the scalability and latency issues in the 5G NR.

\end{abstract}


\IEEEpeerreviewmaketitle

\section{Introduction}

\begin{figure}[h!]
\centering
\includegraphics[width=\linewidth]{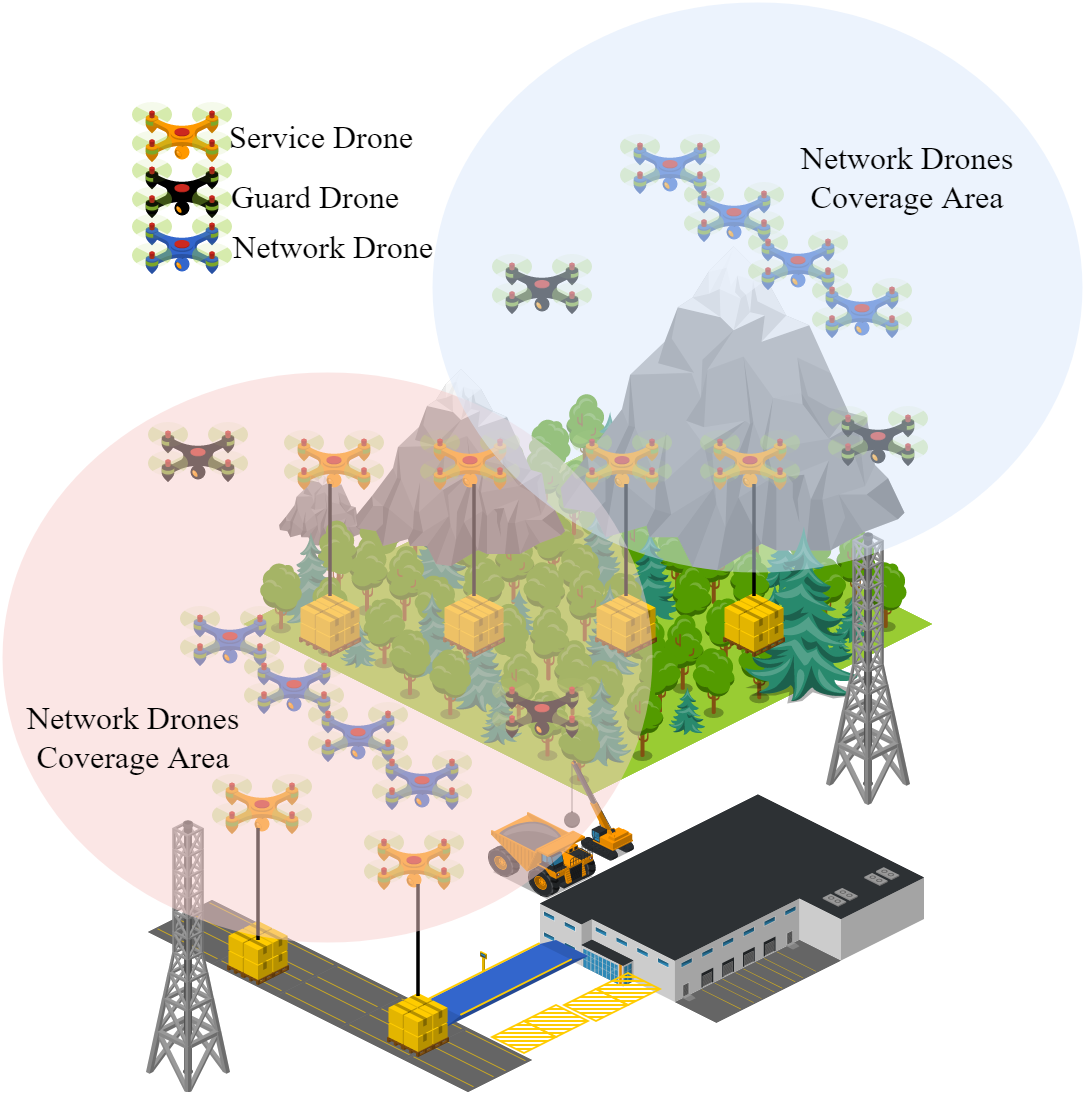}
 \caption{Drone swarm delivering cargo. According to the proposed method, a drone swarm consists of guard, network, and service drones.}
\label{fig:CargoDelivery}
\end{figure}

\lettrine{T}HE USE of drones in the military, agriculture, cargo delivery, and visual shows is increasing swiftly. It is expected that the future of the cargo delivery networks as in Figure \ref{fig:CargoDelivery} will be a combination of transportation and information technology tools \cite{HAPS}. In fact, commercial companies such as Amazon, Alibaba have been initiated to deliver cargo by using their drones for pilot areas. Drone swarms offer better solutions than a single drone in terms of single-point failure, mission area, cost, and performance \cite{intro1}. Drone swarms can fulfill intensive tasks more efficiently than a single drone. If a region for the mission is larger than the coverage of a single drone, drones in the swarm can spread and execute the mission. While fulfilling a task, the drones in the swarm may act autonomously or communicate with each other \cite{intro2}.

The network topology of a drone swarm is not stable due to the limitations of the drones. The limited computing, battery, and storage capabilities are the reason to determine a communication or security protocols for drones. The communication channel for the drone swarms is an open wireless channel and is vulnerable to various attacks such as packet drop attack or routing protocol attacks \cite{intro2}. The actions of each drone in a swarm should always be monitored and any anomaly must be taken care of immediately. Authentication is the first step for the detection of an intruder into the swarm. 

The communication channel between the drones in a swarm is the wireless channel which is vulnerable to eavesdropping \cite{eavesdropping} and hence, the messages between drones should be encrypted to provide confidentiality. Due to the computational efficiency, the symmetric key encryption solutions are the first option to provide confidentiality for the communication channel. A secret key must be known by each party to encrypt the messages via a symmetric key algorithm. The key distribution standard methods becomes a tedious task for such a dynamic swarm where the drones are leaving and participating randomly.

A drone swarm may frequently relocate to accomplish intensive assignments. The connectivity between drones in the swarm and the drone control station is provided by the 5G new radio (NR) network through the base stations (BS) and the core network. While the drone swarms are moving from one region to another, the BS providing radio access for the swarm may have to be shifted. According to the security procedures in 5G NR Release(Rel)-17 \cite{33501}, handover operation between BSs requires data sharing and link update steps. A radio access node on-board of unmanned aerial vehicle (UxNB) may be the serving-BS for the drone swarm. Due to the limitations for unmanned aerial devices (UAV), a target-UxNB can replace a serving-UxNB. 

In this work, we provide an efficient scheme to handle the authentication procedure and key distribution in such situations.The proposed work is an extension of \cite{GA3} where an authentication method is proposed to verify the new drones joining to the swarm. The guard drones were introduced for the authentication of new drones. The group authentication scheme for a new drone and 5G NR UAV authentication were simulated and the results from the simulations were also presented in the previous study.

The drones in the swarm form three sub-groups, which have distinct responsibilities. The guard drones are the drones whose responsibility is to keep track of leaving and joining drones and to authenticate new members in the swarms. The networking operations are executed by the network drones. The main service is provided by the service drones. The proposed method's main contributions can be summarized as;
\begin{itemize}
\item A group key is distributed between drones in the swarm to provide a secure channel for the communication and a solution to share the group key with the new participants is proposed. 

\item Authentication of a new drone participating in the swarm requires two steps if the 5G NR solution is used. The first step is the confirmation of the new drone by the core network and the next step is the authentication via the drone control station. The proposed method offers group-based authentication solution which has better time and communication complexities than 5G NR.

\item During handover operation for the drone swarm, sharing the data for each drone between serving-BS and target-BS may cause latency for the communication. Rather than authentication of each drone one-by-one and sharing information between BSs, the network drones can perform group authentication with the target-BS in the proposed method. 

\item If the service providing BS is a UxNB, the handover steps in 3GPP Rel-17 \cite{33501} cost time and service latency. A group-based handover solution is proposed in the study to provide seamless handover from serving-UxNB to target-UxNB.

\begin{table}[h!]
\caption{Abbreviations}
\label{table:abbreviation}
\centering
 \begin{tabular}{l l}
\hline
\textbf{Abbreviation} & \textbf{Description} \\
 \hline
3GPP & 3rd Generation Partnership Project \\
BS & Base Station \\
UE & User Equipment \\
NR & New Radio\\
UAV & Unmanned Aerial Vehicle \\
UAS & Unmanned Aerial System \\
UxNB & Radio Access Node On Board of UAV \\
UTM & Unmanned Aerial System Traffic Management \\
SUCI & Subscription Concealed Identifier \\
SUPI & Subscription Permanent Identifier \\
UDM & Unified Data Management \\
AMF & Access and Mobility Management Function \\
s-BS & Serving Base Station \\
t-BS & Target Base Station \\
LTE & Long Term Evolution \\
ECDLP & Elliptic Curve Discrete Logarithm Problem \\
MAC & Message Authentication Code \\
ENC & Encryption \\
gNB & Next Generation NodeB \\
\hline
\end{tabular}
\end{table}

This paper is organized as follows. The next section provides an overview of the security aspects of drone swarms in the 3GPP standards and presents brief contribution of related works. In Section III, the preliminaries of our proposed method are laid out in detail. System and threat models are given in Section IV. Our proposed approach for the authentication of an incoming drone and handover via BSs is presented in Section V. The security and performance evaluation is provided in Section VI and Section VII, respectively. The study is completed by a conclusion in Section VIII. We present a list of abbreviations which are used throughout the paper in Table \ref{table:abbreviation}.

\section{Literature Overview and Related Works}
The 3GPP standards and related studies about the security aspects of UAVs are presented in the following part. The details about the current 5G NR solution to authenticate a drone using 3GPP network is also given in this section.

\subsection{UAV Basics in 3GPP}
A UAV and UAV control station constitute the unmanned aerial system (UAS) \cite{22125}. UAVs connect with each other via a 3GPP network. The control station, which controls the UAVs, also uses the 3GPP network. The communication between UAV and control station is within the scope of the 3GPP standards. The communication consists of the command and control directives from the control station to the UAV and uplink or downlink data between UAS components. Unmanned aerial system traffic management (UTM) inside the 3GPP core network provides the services such as authentication of UAVs, enforcing policies to the UAVs, and tracking the location of UAVs.

Regarding the security of the UAS, UTM services should provide confidentiality, integrity, and non-repudiation for the data between the UAVs or from the UAV to the control station. Also, the 3GPP should provide privacy about the identity and location of the UAVs and support regulatory requirements. A UxNB is a radio-access node onboard UAV, which provides service to the UEs. 3GPP can increase the coverage area through the UxNBs. Mobility, airtime, power limitations, and control operations are the main issues while using UAVs as a UxNB.

A BS can ensure services to both the UAVs in the air and the UEs on the ground. The UEs may be responsible for the monitoring and provide real-time video streaming to the main servers. The connectivity for the UEs may be out of service due to the high data transmission by UAVs \cite{22829}. Group communication and authentication should be exploited for the UAVs while scalability and efficiency show up as vital parameters.

\end{itemize}

\subsection{Security Aspects of UAVs in 3GPP Standards}
3GPP TR 33.854 study on security aspects of UAS \cite{333854} is the main document dealing with the security issues of UAVs. The key security issues for UAS and the corresponding solutions are given in the next section. 

\subsubsection{UAS Authentication and Authorization}
Authentication and authorization of the vehicles are the first security issue for the 3GPP Rel-17. Two identification numbers are assigned to a UAV by the UAS service provider and the 3GPP core network. The civil aviation authority level identification number provides the ease of remote identification of a UAV in the air. UAVs utilize the 3GPP identification number when the services provided by the core network are accessed. The authentication of UAVs to provide 3GPP network services is accomplished with two phases. The usual new user equipment (UE) authentication process is performed between UAV and core network in the first phase as shown in Figure \ref{fig:UAVauthentication}. Once the UAV is authenticated by the core network, the UAV control station sends a challenge to the UAV to perform a second authentication. The end-to-end authentication solution between UAV and control station is not covered by the 3GPP standard. The solution is peculiar to the UAS service provider. The steps fo the authentication of a UAV are:

\begin{figure}[h!]
\centering
\includegraphics[width=\linewidth]{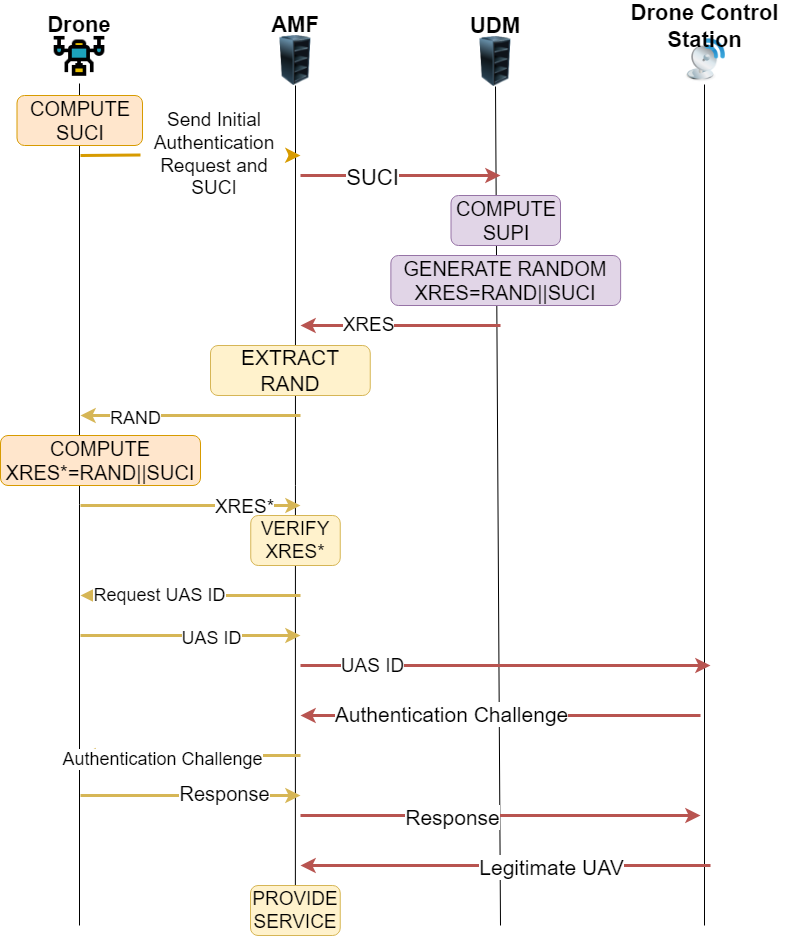}
 \caption{UAV authentication. Authentication of a drone requires two steps according to the 3GPP Rel-17 standard.}
\label{fig:UAVauthentication}
\end{figure}

\begin{enumerate}
	\item The UAV computes the subscription concealed identifier (SUCI) by encrypting the subscription permanent identifier (SUPI) with the base station (BS) public key.
	\item The UAV sends the SUCI to access and mobility management function (AMF).
	\item The AMF shares the SUCI with unified data management (UDM).
	\item The UDM decrypts the SUCI and compares the SUPI with the database.
	\item If the SUPI is valid, the UDM generates a random value and appends the random value with SUCI.
	\item The UDM sends back to AMF the RAND and SUPI.
	\item The AMF extracts the random value and shares it with the UAV.
	\item The UAV appends the SUCI to a random value and sends it back to the AMF.
	\item If the AMF confirms the identity of the UAV, UAS ID is requested from the UAV.
	\item The UAV sends the UAS ID to AMF.
	\item The AMF sends the ID to the control station.
	\item The control station and UAV perform a second authentication.
	\item After control station confirmation, the AMF begins to provide 3GPP service to the UAV.
\end{enumerate}

\subsection{Studies on Security Aspects of Drone Swarms}

The importance of broadcasting for the drone swarm is stated in the study \cite{broadcast}. Following a leader in a swarm is a natural behavior of group communication. The leader prefers to send messages to drones in the swarm as broadcast messages rather than communicating with drones one by one. The authors proposed a broadcast protocol to solve the key distribution issue for the drone swarm. Anytime a new drone participates in the swarm, a new group key is produced and used by the drones. If we take into consideration the dynamic structure of a swarm, all drones will participate in the reconstruction phase of the group key and this will cause too much communication and computational cost for the swarm.

\begin{figure*}[h!]
\centering
\includegraphics[width=\textwidth]{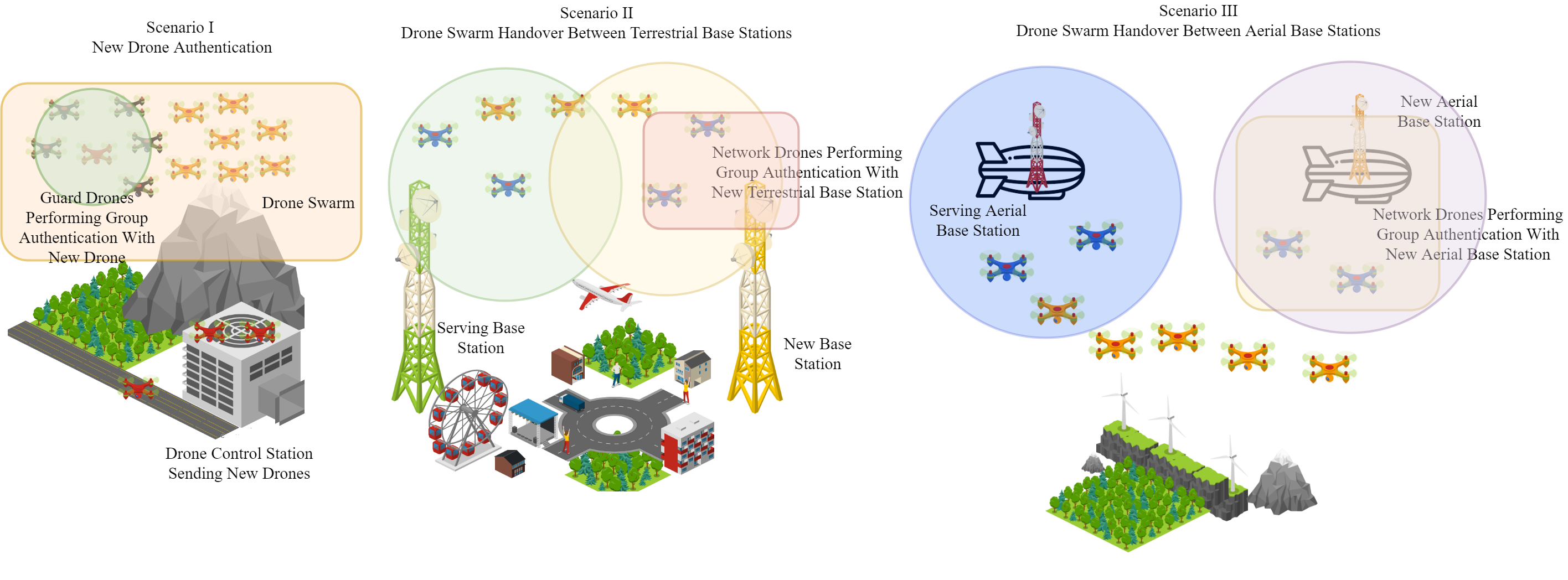}
 \caption{Authentication and handover scenarios for drone swarms.}
\label{fig:scenarios}
\end{figure*}

The secure transmission of aggregated data is accomplished by blockchain in the study \cite{blockchain}. The sensing layer of the internet of things and drones as swarms monitor a predefined region and collect data. The data is sent to the cloud servers via gateways and be part of a blockchain. The security solution is commonly based on the application layer. The security of the sensing and network layers is not taken into consideration.

\section{PRELIMINARIES}
The guard and network drones carry out group authentication solution in \cite{GA} to authenticate a new drone joining the swarm and to reduce handover time from serving to target BS. A polynomial 
\begin{equation}
p(x)=a_0+a_1x+...+a_{m-1}x^{m-1}
\end{equation}
with the most significant power is threshold value $(m)$ is pre-selected according to the group authentication solution. Each member $U_i$ in the group has a unique private key $(p(x_i))$ and a public key $x_i$, $p(x_i)P$. The second pair of the public key is the multiplication of the private key with the pre-selected public point $P$ on an elliptic curve. $a_0$ is the group key that is used to encrypt the messages between members by symmetric key encryption. Point $Q$ is a public point which is the multiplication of group key $(a_0)$ and point $P$. With having valid public and private keys up to the threshold value, a sub-group can perform group authentication.

In the group authentication phase, each member in the group shares its public key with the other members. Once the number of the shares is equal to or greater than the threshold value, each member perform the same computation.
\begin{equation}
c_i=\left(\prod^{m}_{r=1, r\neq i}\dfrac{-x_r}{x_i-x_r})\right)p(x_i)\times P
\end{equation}
If the sum of all $c_i$ is equal to point $Q$, the group authentication is valid.

After the authentication, the member $U_i$ and $U_j$ can generate a secure key with the Diffie-Hellman key exchange protocol. The secure key between the $U_i$ and $U_j$ is $p(x_i)p(x_j)P$. The group members share their private keys with each other to compute the group key. Once the number of the private keys is the threshold value, a group member can compute Lagrange Interpolation
\begin{equation}
group key=a_0=\sum^{m}_{i=1}p(x_i)\prod^{m}_{r=1, r\neq i}\dfrac{-x_r}{x_i-x_r}
\end{equation}
and obtain the group key.

\begin{table}[h!]
\caption{Authentication and Handover Scenarios for Drone Swarms}
\label{table:scenarios}
\centering
 \begin{tabular}{l l}
\hline
\textbf{Scenario} & \textbf{Description} \\
 \hline
Scenario-I & A new drone is requesting to join the swarm. \\
Scenario-II & A drone swarm is performing handover from one\\
&terrestrial BS to the new terrestrial BS. \\
Scenario-III & A UxNB is providing service for the drone swarm\\
&in a rural region. Due to the limitation of the flying time, \\
&the UxNB should be change with a new one. \\
\hline
\end{tabular}
\end{table}

\section{System and Threats Models}
The details of the three system models and the possible attacks for the models are given in the section.

\subsection{System Model}
Three scenarios as shown in Table \ref{table:scenarios} and Figure \ref{fig:scenarios} have been taken into consideration while the framework for the study is prepared. The authentication and handover requirements are the leading points to create the scenarios for the drone swarms.

The first scenario is the authentication of new drones participating in the swarm. A new drone is requesting to join the swarm. The authentication of the new drone should be performed by taking into consideration of latency and sharing a security key. The security key should be used to encrypt the messages between the drones in the swarm after the authentication.

A drone swarm is performing handover from one terrestrial BS to the new terrestrial BS in the second scenario. Authentication of each drone in the swarm by the next BS should be performed by considering the service latency for the drones and the UEs already connected to the new BS.

\begin{figure*}[h!]
\centering
\includegraphics[width=\textwidth]{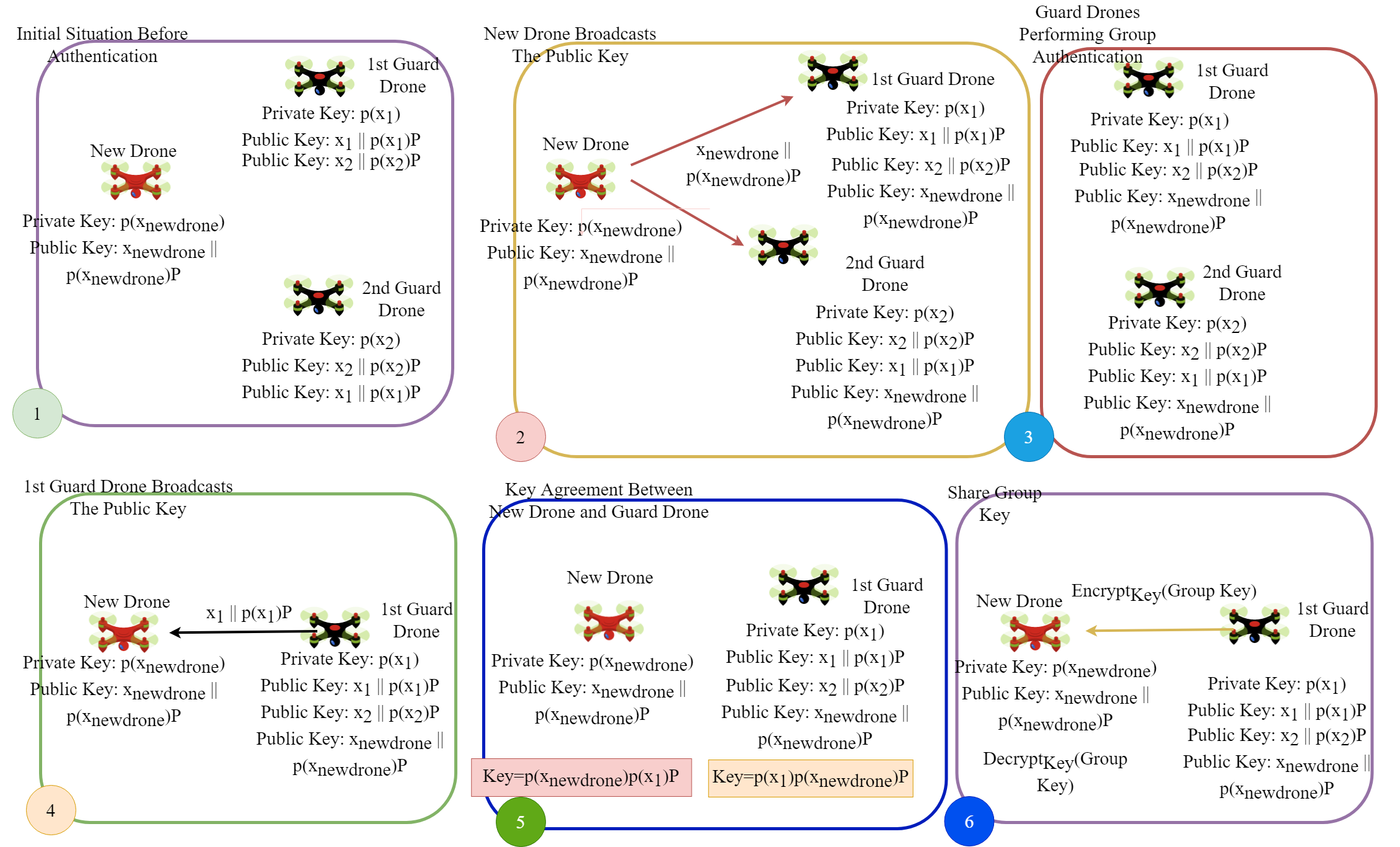}
 \caption{The authentication steps for the new drone.}
\label{fig:step1}
\end{figure*}

In the last scenario, a UxNB is providing service for the drone swarm in a rural region. Due to the limitation of the flying time, the UxNB should be change with a new one. The drone swarm will handover from the serving-UxNB to the target-UxNB.

\subsection{Threat Model}
The communication channel between drones in the air is based on wireless communication. Wireless communication is open to eavesdropping if the messages are not encrypted. The public keys are transferred between drones as plaintext in the group authentication solution. The intruders can capture the traffic and try to obtain useful security information from public keys. Even, the public keys can be exploited for replay attacks.

While a legitimate new drone is performing group authentication with the guard drones, an intruder can capture the public key of the new drone. After a while, the intruder can send a fake drone to the swarm to join an operation. The fake drone sends the public key captured from the previous session.

A fake BS controlled by an attacker can perform group authentication with the drone swarm during a handover process. Once the fake BS is authenticated by drone swarm, the network connectivity is provided via fake BS. All the information produced by drones and transferred to the drone control station may be sniffed by the attackers. The fake BS attacks can be performed with a terrestrial or drone BS. The fake BS attacks can be performed with a terrestrial or drone BS.

\section{Proposed Group Authentication and Handover Solutions for Drone Swarm}
Proposed solutions for the authentication of a new drone joining to the swarm, group handover method for the terrestrial BS handover, and aerial BS handover are described in the following part.

\subsection{Authentication of New Drones by Drone Swarm}
Each drone in the swarm has the group key to encrypt the messages before transmitting them to the other drones. The guard drones should authenticate the new parties willing to be part of the swarm and share the group key with the new party. Before a new drone joins the swarm to provide service and become a service drone, guard drones welcome the new drone as shown in Figure \ref{fig:scenarios} and follows the steps below for authentication:

\begin{figure*}[h!]
\centering
\includegraphics[width=\textwidth]{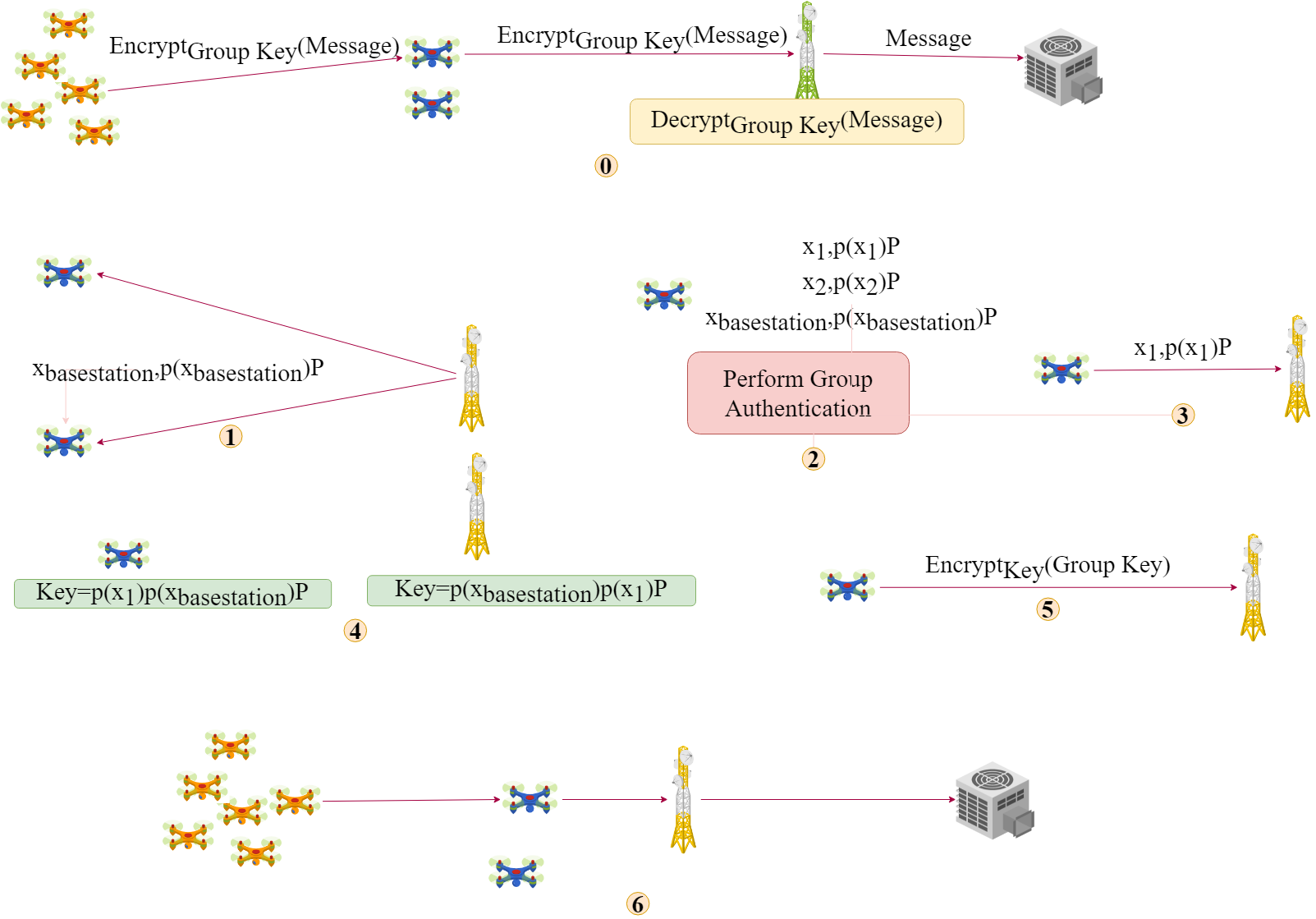}
 \caption{Drone swarm handover between terrestrial BSs.}
\label{fig:step2}
\end{figure*}

\begin{itemize}
\item The control station assigns a private key ($p(x_{new})$) and public key pairs $(x_{new},p(x_{new})P)$ to the new drone as shown in Figure \ref{fig:step1}.
\item The keys are shared with the new drone and the new drone is sent to the drone swarm.
\item The new drone shares the public key pairs with guard drones.
\item The guard drones perform group authentication as in \cite{GA,GA2,GA3}.
\item If the authentication is valid, the pre-defined guard drone perform the key agreement step \cite{GA,GA2,GA3} with the new drone.
\item The group key is encrypted by the agreed key and sent to the new drone.
\item If the authentication is not valid, the new drone is forced to leave the area by guard drones.
\end{itemize}

\subsection{Terrestrial BS Handover}
The territory in which a drone swarm exists may change very speedily as shown in Figure \ref{fig:scenarios}. The movement of numerous drones taking service from one BS to the other area puts burden of the handover loading on the s-BS and t-BS. In our proposed method, the network drones are responsible for the handover process. The number of network drones depends on the threshold value as defined in the group authentication methods at \cite{GA,GA2,GA3}. In order to perform the group authentication as in \cite{GA,GA2,GA3}, the number of group members must be equal or greater than the threshold value. The network drones and BS create a group and perform group authentication. Therefore, the number of network drones must be one shy away from the threshold value. The network drones and t-BS follow the steps below:

\begin{itemize}
\item The network drones share their public key pairs with the t-BS as shown in Figure \ref{fig:step2}.
\item The t-BS performs the group authentication as in \cite{GA,GA2,GA3}.
\item If the authentication is valid, the t-BS begins to provide service to the requests coming from drone swarm.
\end{itemize}

\subsection{Aerial BS Handover}
\begin{figure*}[h!]
\centering
\includegraphics[width=\textwidth]{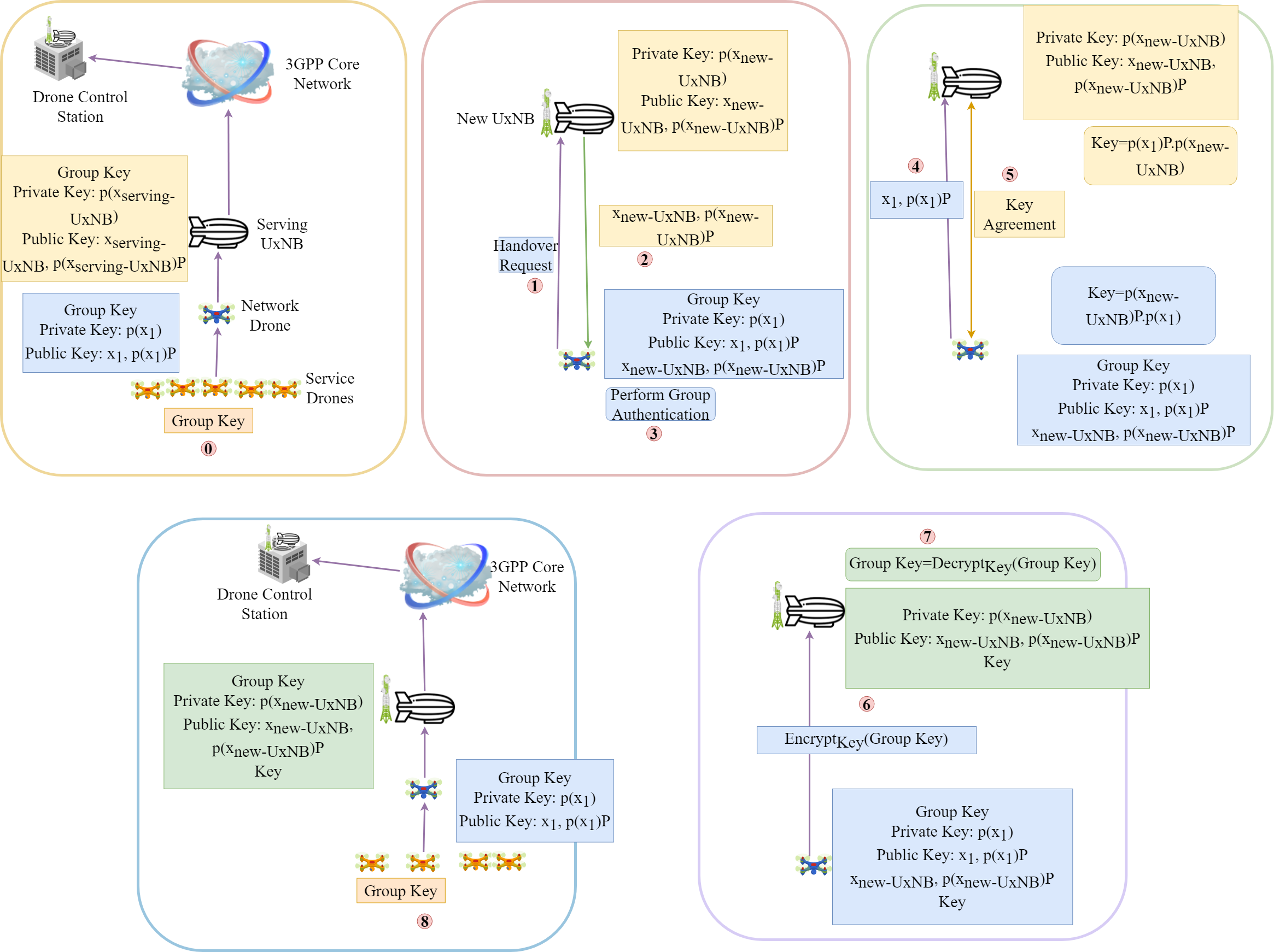}
 \caption{Drone swarm handover between UxNBs}
\label{fig:step3}
\end{figure*}

The connectivity to the core network from the drone swarm may be provided not only by terrestrial BS but also by aerial BS as shown in Figure \ref{fig:step3}. Serving aerial BS may be altered by a new aerial BS due to the limitations of the UAV. Rather than authentication of each drone in the swarm by a new aerial BS, a group authentication between network drones and aerial BS can solve the scalability issues in the handover process. The steps for the group handover are mentioned at below:

\begin{itemize}
\item The new aerial BS shares its public key pairs $(x_{newBS}, p(x_{newBS})P)$ with the network drones.
\item The network drones perform group authentication and verify the new aerial BS.
\end{itemize}

\section{Security Analysis}
The security provided by our proposed method against all possible attack types is explained in this part. %The prevention for an attack is given as an attack and the solution is proved in the prevention for each attack scenario.
\\

\textbf{Attack 1 to Scenario I:} \textit{An intruder can capture public key pairs ($x_i, p(x_i)P$) during a new drone $U_i$ joining operation as explained in Scenario I. A fake drone can perform group authentication with guard drones with the public key pairs captured from the previous session.}

\textit{Prevention.} After group authentication, guard drones, and the new drone perform key agreement process. The new drone $U_i$ uses its private key ($p(x_i)$) in order to compute the encryption key. Due to the absent of a private key, the attackers cannot compute an encryption key to decrypt the group key sent by the guard drones. Without the group key, the fake drone cannot be part of the swarm. {$ \hspace{4 cm} \qed$}  

\textbf{Attack 2 to Scenario II:} \textit{A fake BS with the captured public key from the previous sessions can perform a handover operation with the network drones as explained in Scenario II. If the fake BS may imitate the new target BS for the drone swarm, the traffic from drones to the drone control station can be sniffed by the fake BS.}

\textit{Prevention.} The messages from the drone swarm to the BS are encrypted by the group key. The network drones share the group key with the new BS in encrypted form after the group authentication. The fake BS must have a private key to decrypt the group key. Without the group key, it is not possible to sniff the traffic between drones and the control station. {$ \hspace{7.1 cm} \qed$} 

\textbf{Attack 3 to Any Scenario:} \textit{The attackers can capture the public keys and try to obtain the corresponding private keys from the public ones. The attack can be performed to any scenario since the public key is used for all three scenarios.}

\textit{Prevention.} The public key is the multiplication of the private key and the elliptic curve point $P$. Thanks to the elliptic curve discrete logarithm problem, it is not feasible to have a private key from a public key. {$ \hspace{3.8 cm} \qed$}  

\section{Performance Analysis}
The performance analysis section is divided into two subsections which reflect the analysis for the authentication of drones joining the swarm and the analysis for the handover solutions.

\begin{figure}[h!]
\centering
\includegraphics[width=\linewidth]{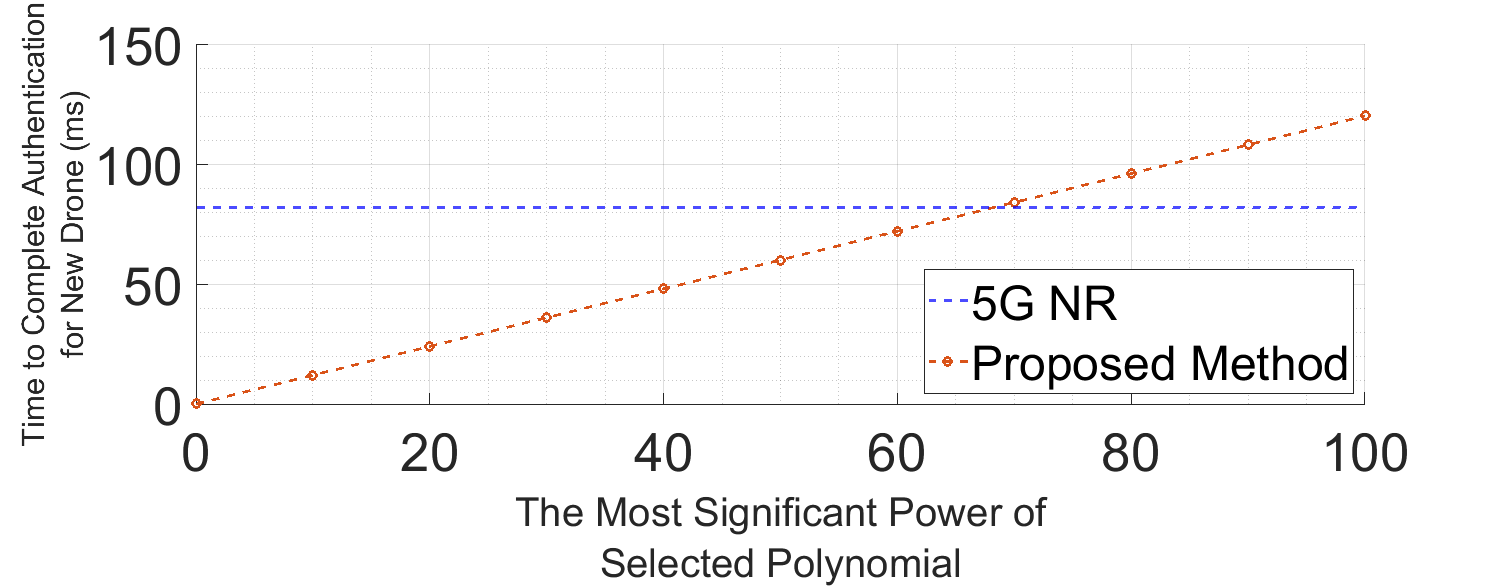}
 \caption{Comparison of time required to authenticate a new drone in 5G NR and our proposed method. The authentication time for a new drone is $82$ \textrm{ms} if 5G NR authentication solution is used, while the authentication time is proportional with the most significant power of selected polynomial in our proposed method. Whenever the most significant power of selected polynomial is less than $70$, the proposed method ensures preferable time complexity than 5G NR.}
\label{fig:performance}
\end{figure}

\subsection{Performance Analysis of Drone Authentication}
The proposed scheme and the UE authentication method used in 5G NR \cite{33501} standards are compared in order to analyze the performance of the study. To authenticate the new drone joining the drone swarm, either the proposed method can be used or the UE's private key (SUPI in 5G NR) is sent to the core network for new UE authentication as specified in Rel 17. The key is confirmed via a query in the database by the core network servers and the new drone is started to be served. In our proposed method, guard drones authenticate the new drone with the group authentication method.

The final result in the performance analysis is to find the time required to verify the identity of the new drone included in the group. This period will be found using both the proposed method and the method in the standards. The simulation is implemented by SimuLTE version 1.2.0 \cite{simulte} library built on top of the Omnet++ package version 5.6.2 and INET framework version 4.2.2. One UE, one BS, and core network servers are deployed to the simulation in order to monitor the time required for network packets transmission. The simulation parameters in the omnet.ini file are left as default. These simulation settings are configured to have the authentication time for the standards. WiFi direct module in SimuLTE is exploited to monitor the time for the communication between UAVs. Two wireless hosts are deployed and a message packet from one host to other is sent to monitor the time. The communication for the authentication in the proposed method is mostly between UAVs. Therefore, the second configuration is about the proposed method.

It has been observed that the time required for a data packet sent by the UE to reach the server in the core network and for the server to send the response back to the UE is approximately $10$ \textrm{ms}. The UE performs one asymmetric encryption operation, which is $100$ \textrm{$\mu$s} \cite{ECP}, to compute SUCI and then sends the SUCI to the core network. The UDM decrypts the SUCI to have SUPI for confirmation. The time for the decryption is approximately $1.5$ \textrm{ms}. Two hashing operations are carried out in the scheme. The UE perform authentication with UDM and drone control station. The time required for the method in the standards is $82$ \textrm{ms} in total due to the $8$ transmisions between UE and core network.

Data exchange time between drones was measured to simulate the proposed method. The wireless direct was carried out with SimuLTE and it was observed that the data transfer took place within $600$ \textrm{$\mu$s} between drones. Each guard drone shares data with the drone as much as the threshold value. Each guard drone also performs an elliptic curve powering operation up to the threshold value. One elliptic curve powering operation is approximately $612$ \textrm{$\mu$s} \cite{ECP}. In total, new drone authentication is performed in approximately $1.2m$ \textrm{ms} ($m$ is the threshold for polynomial) as shown in Figure \ref{fig:performance}. 

\begin{figure}[h!]
\centering
\includegraphics[width=\linewidth]{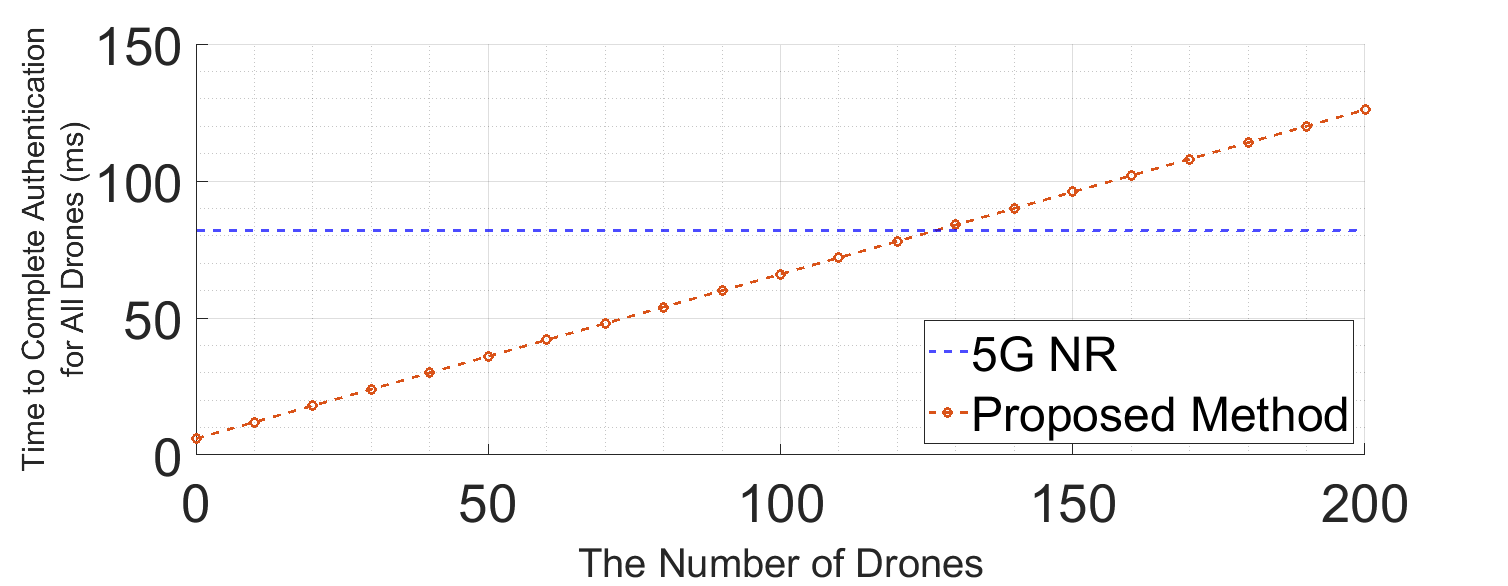}
 \caption{Authentication time for diverse number of drones. If the threshold value is selected as $5$, the proposed method provides less time cost than 5G NR as long as the number of drones is less than $130$.}
\label{fig:performance2}
\end{figure}
\begin{figure}[h!]
\centering
\includegraphics[width=\linewidth]{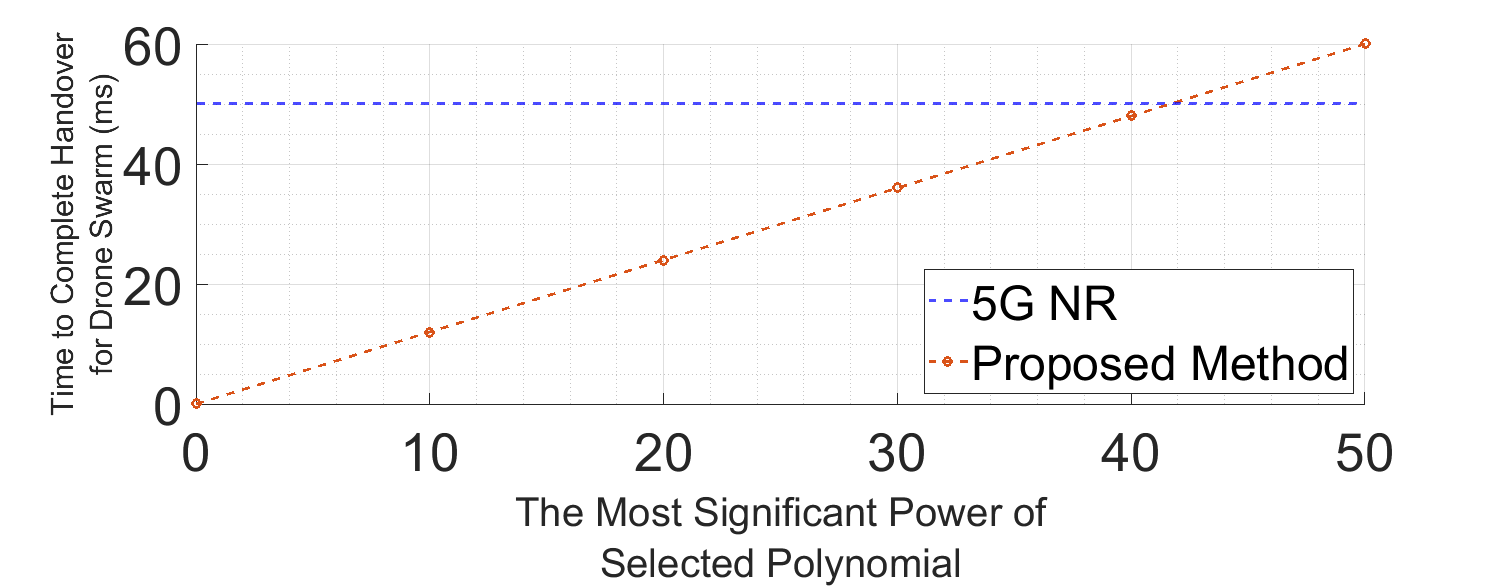}
 \caption{Comparison of handover time in 5G NR and our proposed method. Whenever the most significant power of polynomial is less than $40$, the proposed method ensures preferable time complexity than 5G NR.}
\label{fig:performance3}
\end{figure}
The proposed method provides faster authentication solution than 3GPP Rel-17 according to the simulation results if the most significant power of the polynomial is chosen less than $70$.
The UAV should communicate with the core network two times for the initial authentication according to Rel-16 and Rel-17. 

If it is assumed that $100$ new drones want to participate in the drone swarm, the time required by the core network for authentication is approximately $82$ \textrm{ms} for the 5G NR method as shown in Figure \ref{fig:performance2} if the core network may perform multi-authentication for drones. With the proposed method, the same number of drones can be authenticated as a group. Each party should broadcast its public key pairs, which cost $60$ \textrm{ms} ($600$ \textrm{$\mu$s} x $100$ drones), and one group authentication should be performed, which cost $6$ \textrm{ms} if the threshold value is selected as $5$. Total authentication time is almost $66$ \textrm{ms}, which is extremely less than the 5G NR solution. According to International Mobile Telecommunications-2020 requirements, 5G NR should provide $20$ \textrm{Gbps} data rate and $1$ \textrm{ms} latency \cite{IMT}.

\subsection{Performance Analysis of Handover}
After comparing the results for the new drone authentication, the handover scenario for the drone swarm is simulated in omnetpp. According to the measurement reports from UE, serving-BS decides the handover. If a handover decision is taken, the serving-BS shares the relevant security keys with target-BS. After the data-sharing phase between BSs, the UE de-attaches from serving-BS and attaches to the target-BS. These handover steps as mentioned in 3GPP Rel-17 are simulated in omnetpp to observe the time for the handover in 5G NR. According to the simulation results, the total time for handover operations in 5G NR is $50$ \textrm{ms}.

The network drones in the swarm and target-BS perform a group authentication to complete the handover for the drone swarm. The total time for authentication, which is $1.2m$ \textrm{ms}, depends on the predetermined threshold value. If the threshold value is less than $40$, the proposed handover solution costs less time than the 5G NR as shown in Figure \ref{fig:performance3}.

\section{Conclusion}
In this study, new methods for the authentication of the new drone joining a swarm and handover scenarios for the drone swarms are presented. Rather than sending authentication requests to the 3GPP core network and drone control station for each drone joining the swarm, a group authentication method is proposed in the study. According to the simulation results, the proposed method costs less time than 5G NR.

In addition to the authentication of new drones, another focus of this study is handover process of drone swarms. On behalf of the drones in the swarm, the network drones perform group authentication with target-BS during the handover process. The simulation results indicate that the time required for the handover in the proposed method is less than the 5G NR handover process if the threshold value is selected less than $40$. Moreover, the handover solution for the mobile BSs is also proposed in the study.

\begin{IEEEbiography}[{\includegraphics[width=1in,height=1.25in,clip,keepaspectratio]{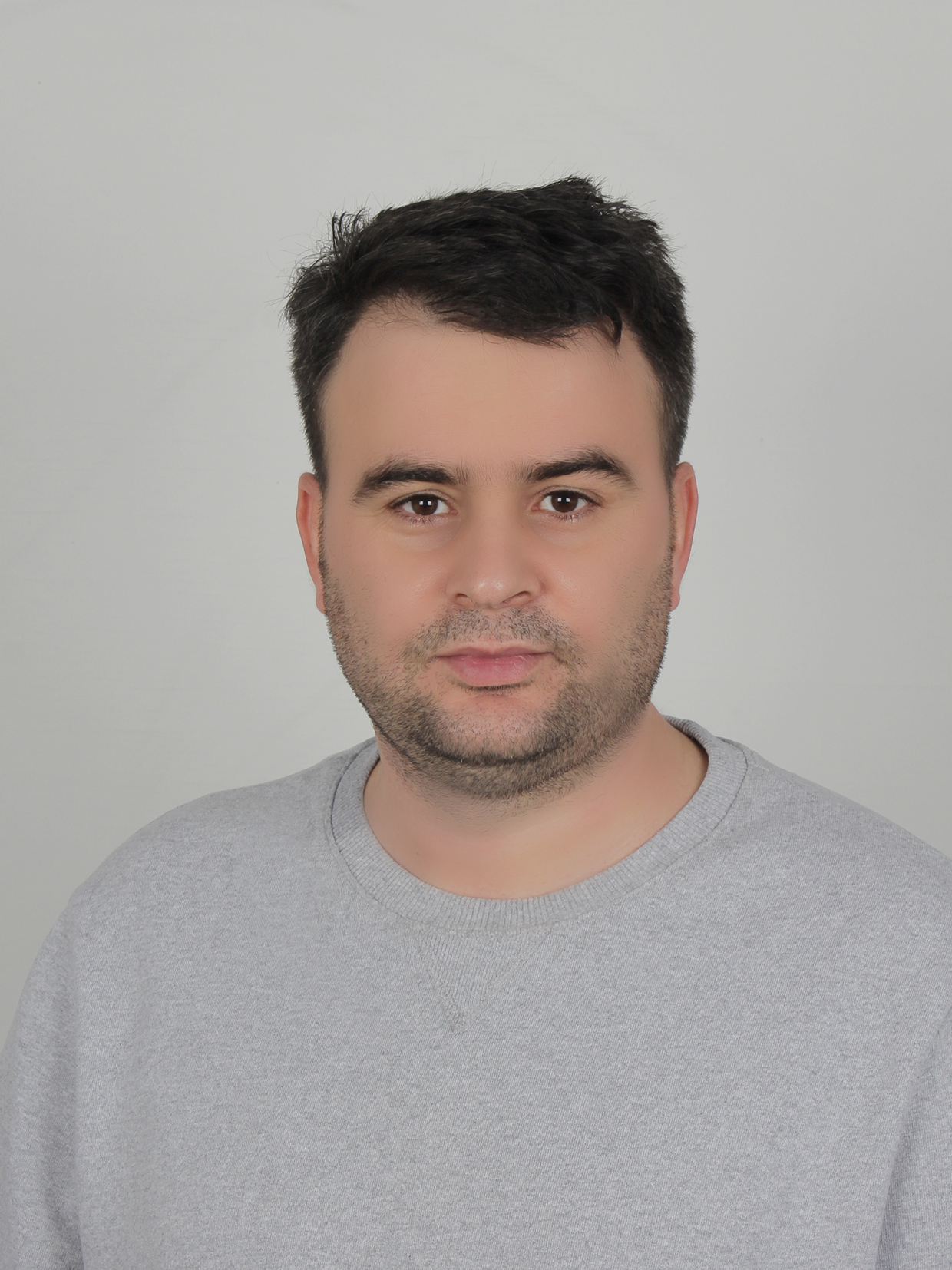}}]{Yucel Aydin}
(aydinyuc@itu.edu.tr) received the M.Sc. degree in cyber security engineering and cryptography program from Informatics Institute, Istanbul Technical University, Istanbul, Turkey, in 2017, where he is currently pursuing the Ph.D. degree.

His research interests are network security, cryptography and computer security.
\end{IEEEbiography}

\begin{IEEEbiography}[{\includegraphics[width=1in,height=1.25in,clip,keepaspectratio]{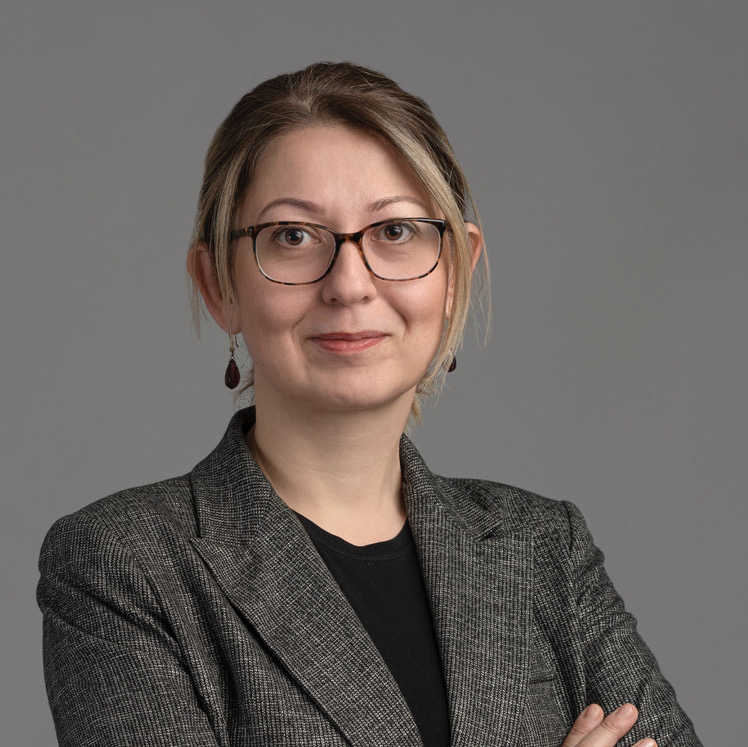}}]{Gunes Karabulut Kurt}
[StM’00, M’06, SM’15] (gunes.kurt@polymtl.ca) is currently an Associate Professor of Electrical Engineering at Polytechnique Montréal, Montreal, QC, Canada. She received the B.S. degree with high honors in electronics and electrical engineering from the Bogazici University, Istanbul, Turkey, in 2000 and the M.A.Sc. and the Ph.D. degrees in electrical engineering from the University of Ottawa, ON, Canada, in 2002 and 2006, respectively. From 2000 to 2005, she was a Research Assistant at the University of Ottawa. 

Between 2005 and 2006, Gunes was with TenXc Wireless, Canada. From 2006 to 2008, she was with Edgewater Computer Systems Inc., Canada. From 2008 to 2010, she was with Turkcell Research and Development Applied Research and Technology, Istanbul. Gunes was with Istanbul Technical University from 2010 to 2021. She is a Marie Curie Fellow and has received the Turkish Academy of Sciences Outstanding Young Scientist (TÜBA-GEBIP) Award in 2019. She is an Adjunct Research Professor at Carleton University. She is also currently serving as an Associate Technical Editor (ATE) of the \textit{IEEE Communications Magazine} and a member of the IEEE WCNC Steering Board.
\end{IEEEbiography}

\begin{IEEEbiography}[{\includegraphics[width=1in,height=1.25in,clip,keepaspectratio]{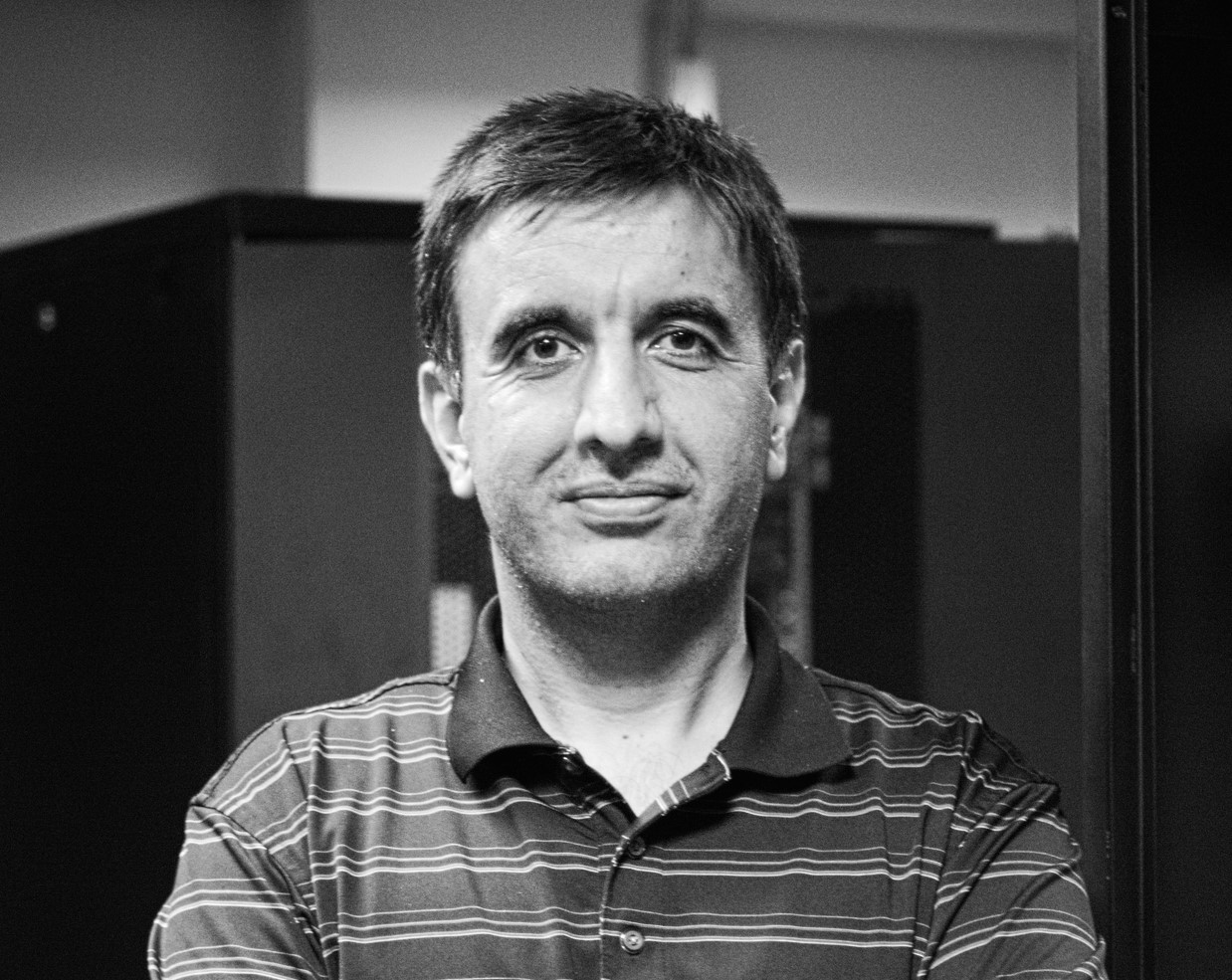}}]{Enver Ozdemir}
(ozdemiren@itu.edu.tr) received a Ph.D. degree in mathematics from the University of Maryland, College Park, MD, USA, in 2009.
He is currently an Associate Professor at Informatics Institute, Istanbul Technical University, Istanbul, Turkey. He was a member of the Coding Theory and Cryptography Research Group, Nanyang Technological University, Singapore from 2010 to 2014. He is also the deputy director of National Center for High Performance computing. His research interests include cryptography, computational number theory and network security.
\end{IEEEbiography}

\begin{IEEEbiography}[{\includegraphics[width=1in,height=1.25in,clip,keepaspectratio]{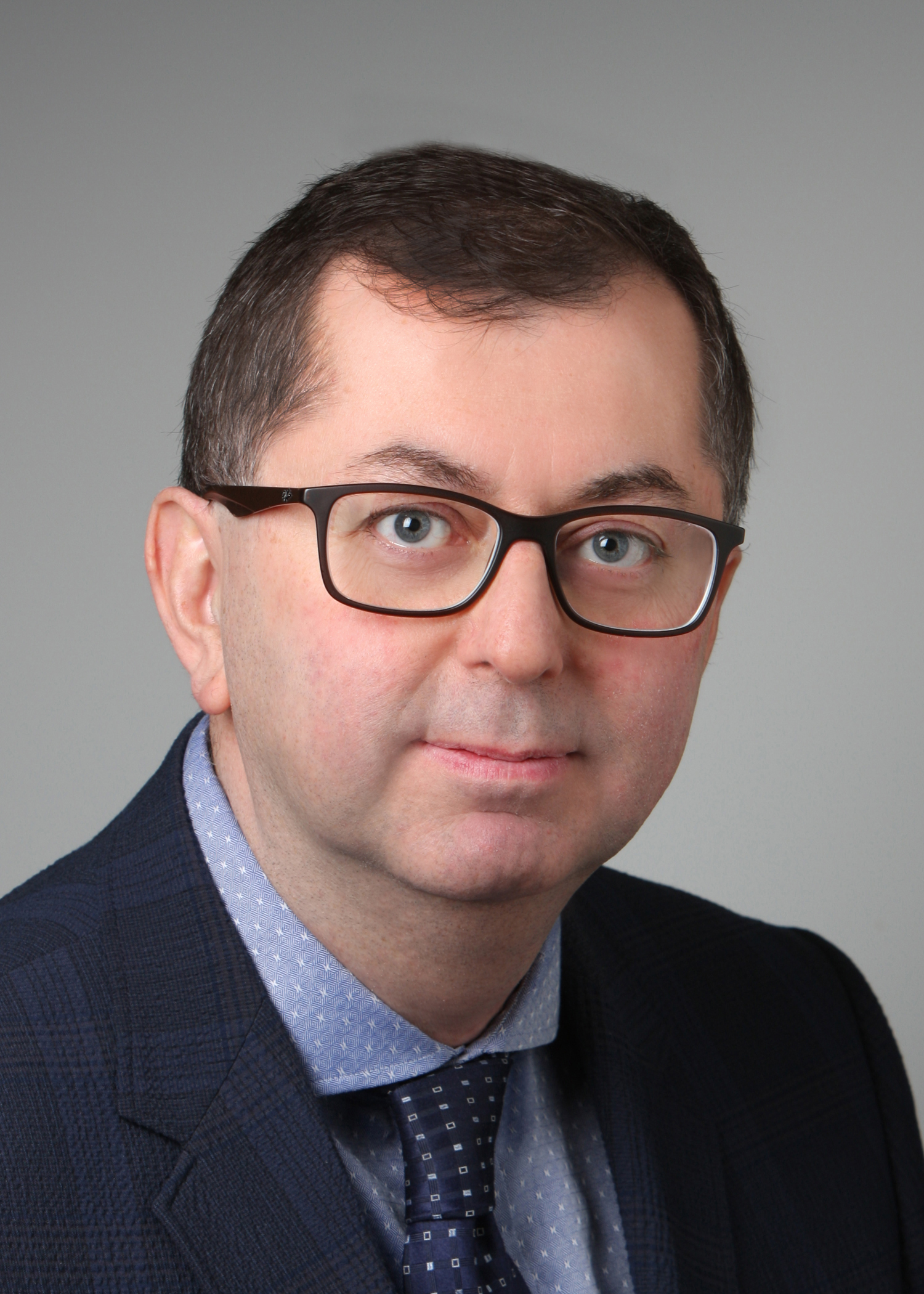}}]{Halim Yanikomeroglu}
[F] (halim@sce.carleton.ca) is a Professor in the Department of Systems and Computer Engineering at Carleton University, Ottawa, Canada. His primary research domain is wireless communications and networks. His research group has made substantial contributions to 4G and 5G wireless technologies. During 2012-2016, he led one of the largest academic-industrial collaborative research programs on pre-standards 5G wireless. In Summer 2019, he started a new large-scale project on the 6G non-terrestrial networks. His extensive collaboration with industry resulted in 39 granted patents. 

He has formally supervised or hosted at Carleton a total of 135 postgraduate researchers in all levels (PhD \& MASc students, PDFs, and professors). He has coauthored IEEE papers with faculty members in 80+ universities in 25 countries and industry researchers in 10 countries. He is a Fellow of IEEE, EIC (Engineering Institute of Canada), and CAE (Canadian Academy of Engineering), and a Distinguished Speaker for both IEEE Communications Society and IEEE Vehicular Technology Society. 

He is currently serving as the Chair of the IEEE WCNC (Wireless Communications and Networking Conference) Steering Committee. He was the Technical Program Chair/Co-Chair of WCNC 2004 (Atlanta), WCNC 2008 (Las Vegas), and WCNC 2014 (Istanbul). He was the General Chair of IEEE VTC 2010-Fall (Ottawa) and VTC 2017-Fall (Toronto). He also served as the Chair of the IEEE’s Technical Committee on Personal Communications. Dr. Yanikomeroglu received several awards for his research, teaching, and service, including the IEEE Communications Society Wireless Communications Technical Committee Recognition Award in 2018, IEEE Vehicular Technology Society Stuart Meyer Memorial Award in 2020, and IEEE Communications Society Fred W. Ellersick Prize in 2021.
\end{IEEEbiography}

\end{document}